\documentclass[aps,prx,twocolumn,superscriptaddress,showpacs,floatfix]{revtex4-1}

\usepackage{amsmath,amssymb,graphicx}

\usepackage[utf8]{inputenc}
\usepackage[T1]{fontenc}
\usepackage{xcolor}
\usepackage{float}

\IfFileExists{newtxtext.sty}
{\usepackage{newtxtext,newtxmath}}
{\IfFileExists{stix.sty}
	{\usepackage{stix}}
	{\IfFileExists{mathptmx.sty}
		{\usepackage{mathptmx}}{} } }

\usepackage{textcomp}

\usepackage{bm}

\IfFileExists{siunitx.sty}{\usepackage{booktabs,siunitx}}{}

\usepackage{color}
\definecolor{LinkColor}{rgb}{0.256,0.439,0.588}
\usepackage{hyperref}

\usepackage{pifont}

\begin{document}

\title{Emergent O(4) symmetry at the phase transition from  plaquette-singlet to
antiferromagnetic order in 
quasi-two-dimensional quantum magnets}

\author{Guangyu Sun}
\affiliation{Beijing National Laboratory for Condensed Matter Physics and
Institute of Physics, Chinese Academy of Sciences, Beijing 100190, China} 
\affiliation{School of Physical Sciences, University of Chinese Academy of Sciences, Beijing 100190, China}

\author{Nvsen Ma}
\affiliation{School of Physics, Key Laboratory of Micro-Nano Measurement-Manipulation and Physics, Beihang University, Beijing 100191, China}
\affiliation{Beijing National Laboratory for Condensed Matter Physics and
Institute of Physics, Chinese Academy of Sciences, Beijing 100190, China} 

\author{Bowen Zhao}
\affiliation{Department of Physics, Boston University, 590 Commonwealth Avenue, Boston, Massachusetts 02215, USA}

\author{Anders W. Sandvik}
\email{sandvik@bu.edu}
\affiliation{Department of Physics, Boston University, 590 Commonwealth Avenue, Boston, Massachusetts 02215, USA}
\affiliation{Beijing National Laboratory for Condensed Matter Physics and
Institute of Physics, Chinese Academy of Sciences, Beijing 100190, China} 

\author{Zi Yang Meng}
\email{zymeng@iphy.ac.cn}
\affiliation{Beijing National Laboratory for Condensed Matter Physics and
Institute of Physics, Chinese Academy of Sciences, Beijing 100190, China}
\affiliation{Department of Physics and HKU-UCAS Joint Institute of Theoretical and Computational Physics,
The University of Hong Kong, Pokfulam Road, Hong Kong}
%\affiliation{Songshan Lake Materials Laboratory, Dongguan, Guangdong 523808, China}

\begin{abstract}
Recent experiments [J.~Guo et al., Phys.~Rev.~Lett.~{\bf 124}, 206602 (2020)] on thermodynamic properties of the frustrated layered quantum magnet 
SrCu$_2$(BO$_3$)$_2$---the Shastry-Sutherland material---have provided strong evidence for a low-temperature phase transition between plaquette-singlet and
antiferromagnetic order as a function of pressure. Further motivated by the recently discovered unusual first-order quantum phase transition with 
an apparent emergent O(4) symmetry of the antiferromagnetic and plaquette-singlet order parameters in a two-dimensional ``checkerboard J-Q'' quantum 
spin model [B.~Zhao et al., Nat.~Phys.~{\bf 15}, 678 (2019)], we here study the same model in the presence of weak inter-layer couplings. Our focus is on the evolution of the emergent symmetry as the system crosses over from two to three dimensions and the phase transition extends from strictly
zero temperature in two dimensions up to finite temperature as expected in SrCu$_2$(BO$_3$)$_2$. Using quantum Monte Carlo simulations, we map out the phase 
boundaries of the plaquette-singlet and antiferromagnetic phases, with particular focus on the triple point where these two ordered phases meet the
paramagnetic phase for given strength of the inter-layer coupling. All transitions are first-order in the neighborhood of the triple point.
We show that the emergent O(4) symmetry of the coexistence state breaks down clearly when the interlayer coupling becomes sufficiently large, but 
for a weak coupling, of the magnitude expected experimentally, the enlarged symmetry can still be observed at the triple point up to significant length 
scales. Thus, it is likely that the plaquette-singlet to antiferromagnetic transition in SrCu$_2$(BO$_3$)$_2$ exhibits remnants of emergent O(4)
symmetry, which should be observable due to additional weakly gapped Goldstone modes.



\end{abstract}

\date{\today}

\maketitle

\section{Introduction}
\label{sec:intro}

In recent years, emergent symmetries in quantum magnets hosting phase transitions between different symmetry breaking ground states have been studied actively.
In classical systems with phases breaking O($N$) symmetry with different numbers of spin components $N$, several theoretical works have addressed the possibility of
O$(N_1+N_2)$ symmetry when O$(N_1)$ and O$(N_2)$ phases meet \cite{Calabrese2003,Eichhorn2013}. In quantum many-body systems, a possible emergent SO(5) symmetry 
was intensely studied in the context of the transition between an O($3$) antiferromagnet and a d-wave superconductor \cite{SCZhang1999,JPHU2001}. Emergent continuous 
symmetry is also an integral aspect of the two-dimensional (2D) deconfined quantum-critical point (DQCP) \cite{deconfine1,deconfine2}, where a dimerized valence-bond
solid (VBS) with $Z_4$ symmetry breaking attains U(1) symmetry upon approach to the DQCP \cite{levinsenthil,Sandvik2007,Jiang2008,Lou2009} and there may possibly
be emergent  SO(5) symmetry when this order parameter combines with the O(3) antiferromagnetic (AFM) order parameter
\cite{senthilfisher,Nahum2015b,Suwa2016,Wang2017,Gazit2018,Sreejith2019,HongYao2019,Sato2020}.
Similarly, a planar antiferromagnet may develop O(4) symmetry at its DQCP \cite{YQQin2017,NvsenMa2018a,ma2019role} when the U(1) magnetic order
combines with the VBS state hosting emergent U(1) symmetry. While the ultimate nature of the 
DQCP is still controversial---truly continuous or weakly first-order---it has by now been established in many numerical studies that spinon deconfinement 
and the proposed emergent symmetries can exist on sufficiently large length scales (hundreds of lattice spacings) for the DQCP phenomenology to apply
\cite{Sandvik2007,Jiang2008,Lou2009,Melko2008,Tang2013,Block2013,Harada2013,Nahum2015a,Nahum2015b,Pujari2015,Shao2016,YQQin2017,XFZhang2017,NvsenMa2018a,ma2019role,SandvikCPL2020,Bowen2020CPB}. The alternative scenario of a weakly first-order \cite{Jiang2008,Chen2013} transition is also interesting in that it may correspond to a non-unitary conformal field theory, with a DQCP outside the accessible model space, e.g., in space-time dimensionality slightly less than $2+1$
\cite{Wang2017,RMa2020,Nahum2020}.

Following the recent surprising discovery of emergent O(4) symmetry of the coexistence state at a first-order quantum phase transition in a 
2D ``checker-board J-Q'' (CBJQ) model with O(3) AFM and two-fold degenerate plaquette-singlet (PS) ground states \cite{BWZhao2018}, we here study this type of 
transition in the presence of weak three-dimensional (3D) interlayer couplings. Such couplings are unavoidable in experimental quasi-2D quantum magnets
that may host DQCP-like PS--AFM transitions---specifically the promising case of the Shastry-Sutherland (SS) compound SrCu$_2$(BO$_3$)$_2$ \cite{Zayed17,Lee2019,JingGuo2020,Bettler2020}. We focus our computational model study on the persistence and eventual break-down of the O(4) symmetry of the coexistence state as the interlayer
coupling is increased from zero and the phase transition extends from the ground state to finite temperature. This phenomenon has immediate relevance to the
decades of works devoted to the SS material, where the PS--AFM transition is still under active pursuit and the efforts have been further reinvigorated by
recent experimental progress at the high pressures and low temperatures where the transition should occur \cite{Zayed17,JingGuo2020,Bettler2020}. We continue 
in this introductory section with a summary of the recent experimental and theoretical developments that motivate our quantum Monte Carlo (QMC) study of 
PS--AFM phase transitions in the quasi-2D geometry.

\subsection{The Shastry-Sutherland material}

The currently most promising system for experimentally investigating DQCP related phenomena is the SS material SrCu$_2$(BO$_3$)$_2$,
a layered $S=1/2$ quantum magnet where the interactions between the unpaired spins on Cu sites within the layers are well described by the 2D frustrated 
SS model \cite{SS1981}, which comprises two Heisenberg couplings; inter-dimer $J$ and intra-dimer $J'$. The SS model hosts three ground state phases
versus the coupling 
ratio $\alpha=J/J'$; dimer-singlet (DS), PS, and AFM. The DS ground state is a unique exact product state of singlets on each of the $J'$ 
bonds, while the PS phase is two-fold degenerate, corresponding to two possible ways of forming four-spin singlets (strictly speaking alternating
higher and lower singlet density) on the ``empty'' plaquettes (i.e., those without $J'$ couplings). The AFM phase is akin to that in the conventional
square-lattice $S=1/2$ Heisenberg model, to which the SS model reduces in the limit $\alpha \to \infty$. 

At ambient pressure SrCu$_2$(BO$_3$)$_2$ is in the DS phase, as has been shown in many different experiments. Early on magnetic susceptibility, Cu
nuclear quadrupole resonance, and high-field magnetization measurements \cite{Kageyama1999,Miyahara1999} established a gapped state corresponding to SS couplings
inside the DS phase but rather close to the PS boundary. More detailed studies followed using NMR \cite{Haravifard2016,Waki2007}, X-ray diffraction 
\cite{Haravifard2012,LOA2005980}, and electron spin resonance \cite{Sakurai_2009}. The other two expected phases have also recently been 
identified under high pressure $P$ \cite{Zayed17,JingGuo2020,Bettler2020}, where the SS couplings $J$ and $J'$ change significantly and unequally 
with $P$, so that the ratio $\alpha$ spans the entire range of the three expected phases for $P$ up to 4 GPa. 

Inelastic neutron scattering experiments detected an excitation mode argued to originate from a PS state at $P=2.15$ GPa \cite{Zayed17}. Subsequently, a phase 
transition at temperature
$T \approx 2$ K was detected in heat capacity measurements between $1.7$ and $2.4$ GPa \cite{JingGuo2020}. It was also 
shown by these experiments that the spin gap changes discontinuously between two different non-zero values at $P=1.7$ GPa, in accord with the first-order 
(level crossing) transition between the DS and PS phases of the SS model. Furthermore, at higher pressures signatures in the heat  
capacity indicate a transition into a gapless phase, most likely the SS AFM phase, with the transition temperature ranging from $2.5$ K at $3$ GPa to 
about $4$ K at $4$ GPa \cite{JingGuo2020}. Above $4$ GPa the system undergoes a structural transition, after which the SS description is no longer valid. 

The earlier neutron scattering experiments had also detected AFM order extending up about $120$ K around $4$ GPa, and it was argued that this was
the SS AFM phase \cite{Zayed17}. However, this interpretation is implausible because of the large mismatch between the high transition temperature, given
the values of the couplings and the high level of geometric frustration (which lowers the effective energy scale), and the much lower temperature scale
of the PS phase. Most likely, the high-temperature AFM phase above $4$ GPa has a different origin related to the structural transition, as does a still
unknown gapless phase detected in the same pressure region below $9$ K \cite{JingGuo2020}. A low-temperature AFM phase starting above 2.4 GPa is also consistent
with the nature of the short-range spin correlations  detected using Raman spectroscopy slightly above the transition temperature \cite{Bettler2020}.

The demonstration of a low-temperature AFM phase between $3$ and $4$ GPa has solidified the expectations from the SS model of a direct PS--AFM quantum 
phase transition in SrCu$_2$(BO$_3$)$_2$, though it occurs below the lowest accessible temperature, 1.5 K, in the recent heat capacity experiments between 
$2.4$ and $3$ GPa \cite{JingGuo2020}. In Ref.~\onlinecite{Jimenez2020} results were reported at lower temperatures but only up to 2.65 GPa and still with
no sign of the PS--AFM transition (though the AFM phase was detected at the highest pressures when a high magnetic field was applied). Neutron scattering 
experiments at these pressures are very challenging below $4$ K. Though the nature 
of the PS phase in SrCu$_2$(BO$_3$)$_2$ is still under investigation, in both scenarios of full-plaquette and empty-plaquette PS state \cite{Boos2019,Jimenez2020} 
there is spontaneous breaking of a two-fold symmetry (phonon assisted or purely spin-driven). Thus, in either case there should be a direct transition 
between a $Z_2$ symmetry-breaking singlet phase and an O(3)-breaking AFM phase. Efforts to actually detect this transition are driven by the prospects of 
identifying the first experimental realization of the DQCP phenomenon in a quantum magnet.  

\subsection{Weak first-order transitions and emergent symmetries}
\label{sec:intro_emergent}
  
On the theoretical side, direct QMC simulations in the entire  parameter range of the SS model are hampered by the sign problem associated with the geometrically
frustrated Heisenberg couplings. Changing the simulation basis from the standard individual spin-$z$ components to the singlet-triplet states on the SS dimers
alleviates the sign problem, and some QMC results for the heat capacity have been obtained in the DS phase~\cite{Wessel2018,Wietek2019}. The PS phase and its
transition into the AFM phase are still beyond QMC simulations. Impressive progress has been made with alternative techniques such as the density matrix
renormalization group (DMRG) method \cite{dmrg01,Schollwock2011} and tensor network states \cite{Orus2014}, and the locations of the DS--PS and PS--AFM quantum
phase transitions obtained from such calculations with the SS model are now considered reliable \cite{sstensor,Boos2019,Lee2019}. Quantitatively establishing the
nature of the transitions is still challenging, however. 

Finite-temperature properties of frustrated systems can to some extent be studied using exact
diagonalization and finite-temperature Lanczos methods \cite{Prelovsek2018}, and progress has also been made recently with extensions of the DMRG method
\cite{Chen2018c,Chen2019}. In the case of the SS model, c alculations can characterize, e.g., the dominant broad peak in the heat capacity
\cite{JingGuo2020,Jimenez2020,Shimokawa2020} (which recently was shown to reflect an analogue of a gas--liquid critical point in SrCu$_2$(BO$_3$)$_2$ at
$P\approx 2$ GPa, $T \approx 4$ K), but cannot resolve the lower-temperature peaks developing at the PS and AFM ordering transitions.

Some ground state calculations indicated a weak first-order PS--AFM transition \cite{sstensor} in the SS model, while other works have suggested a continuous DQCP
transition \cite{Lee2019}. In the latter case, the system sizes used in DMRG calculations were unavoidably
rather small, and weak first-order behavior may still emerge for larger 
systems. From the field theory side, the expectation is that the U(1) symmetry of the singlet phase associated with the DQCP phenomenon cannot emerge from the 
$Z_2$ order parameter of the two-fold degenerate PS phase \cite{Block2013,Nakayama2016}, while that is possible with the $Z_4$ columnar VBS order parameter that has 
been studied with 2D J-Q models \cite{Sandvik2007,Lou2009,Melko2008,Jiang2008,Tang2013,Block2013,Harada2013,Chen2013,Pujari2015,Shao2016,SandvikCPL2020} 
and 3D classical loop models \cite{Nahum2015a,Nahum2015b} (and we note that the case of a $Z_3$ VBS order parameter on the honeycomb lattice is unsettled as
regards the emergent symmetry \cite{Pujari2015,Nakayama2016}). Nevertheless, if the correlation length at a weak first-order PS--AFM transition is sufficiently
large, it may still be possible to observe remnants of the higher symmetries and other phenomena associated with spinon deconfinement. Indeed, the excitations
identified in the SS model by the DMRG calculations for systems with up to hundreds of spins are consistent with the DQCP scenario \cite{Lee2019}.

Given that direct numerical studies of the frustrated SS model are still challenging, it is also useful to investigate alternative ``designer hamiltonian'' with
the  same symmetries and ground-state phases, and which are tailored to be amenable to, in particular, large-scale unbiased QMC simulations \cite{Kaul2013a}.  
The CBJQ model was introduced in this context in order to study the quantum phase transition between a $Z_2$ PS phase and an O(3) AFM phase \cite{BWZhao2018}.

The original square-lattice $J$-$Q$ model \cite{Sandvik2007} combines the $S=1/2$ Heisenberg model with four-spin plaquette terms of strength $Q$ that are not 
frustrated in the conventional sense, yet compete against the AFM order by inducing local correlated singlets. For sufficiently large $Q/J$, these interactions 
drive the system into a columnar VBS phase. The model is emendable to QMC simulations and has been one of the key computational frameworks within which to investigate 
the DQCP phenomenon. The simplest $Q$ term is a product of singlet projectors on two adjacent bonds, and generalized interactions formed from more than two singlet 
projectors have also been extensively studied \cite{Lou2009,Sen2010,Takahashi2020}. In the CBJQ model, the four-spin (two singlet projectors) terms are only included 
on half of the square-lattice plaquettes, forming a checker-board pattern. This arrangement reduces the lattice symmetries and allows for a $Z_2$ breaking PS 
state similar to that in the SS model. 

QMC studies of the CBJQ model revealed a clearly first-order AFM--PS quantum phase transition \cite{BWZhao2018}. However, the coexistence state at the transition
is of an unusual kind, where no tunneling barriers between the PS and AFM phases were detected and the fluctuations appear to obey O(4) symmetry. The symmetry
was characterized using the probability distribution of the combined vector order parameter $(m_x,m_y,m_z,d)$, where the first three components are those of 
the O(3) AFM order parameter and $d$ is the scalar PS order parameter. Moreover, the PS phase extends to finite temperature with a critical temperature
depending on the distance of the coupling ratio $g$ from the $T=0$ transition point $g_c$ according to a logarithmic form, as expected for an O($N$) 
model with Ising-like anisotropy (where the $d$ corresponds to the Ising component of the order parameter) \cite{Irkhin1998}. Thus, while it is not known 
whether the O(4) symmetry is truly manifested asymptotically, it exist on length scales large enough, at least $10^2$ lattice spacings based on the system 
sizes studied, to have consequences for the phase diagram and the low-energy excitations. A similar phase transition was detected in a deformed 3D classical
loop model \cite{Serna2019}, and later a $J$-$Q$ model with $Z_4$ symmetry breaking PS was shown to host a first-order transition with emergent SO(5)
symmetry \cite{Takahashi2020}. 

The SS model has the same order parameter symmetries as the CBJQ model, and it is likely that its PS--AFM transition also hosts emergent O(4) symmetry. 
Moreover, since SrCu$_2$(BO$_3$)$_2$ should also undergo a PS--AFM transition, somewhere between $P= 2.6$ GPa and $3$ GPa \cite{JingGuo2020,Jimenez2020}, 
there are  now unique opportunities to study an emergent symmetry experimentally. The calculations to be presented in this paper will address the feasibility of 
the emergent symmetry surviving when the 3D couplings in the material are taken into account.

\begin{figure}[t]
\centering
\includegraphics[width=84mm]{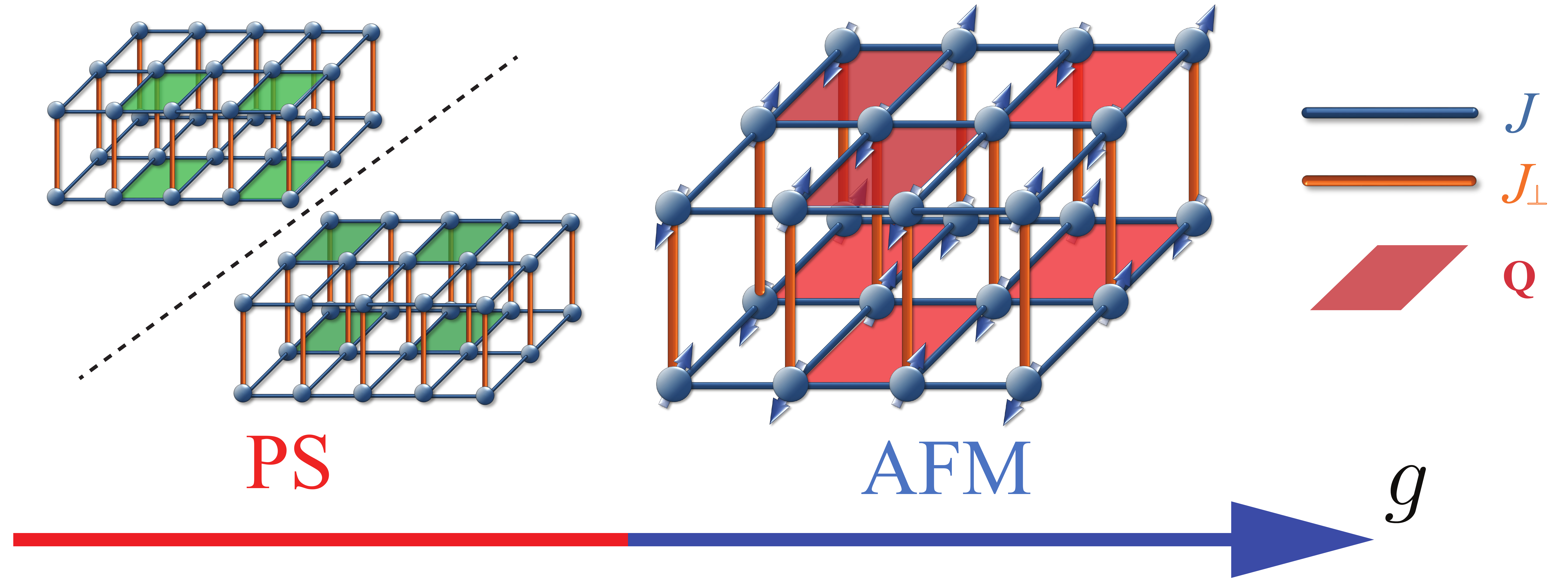}
\caption{3D CBJQ model and its symmetry-breaking ground states versus the coupling ratio $g={J}/{(J+Q)}$. Here $J$ and $Q$  are, respectively,
the intralayer Heisenberg interactions and four-spin plaquette interactions, the latter indicated by the red plaquettes in the illustration to the right. 
In the PS phase, the singlet density on these plaquettes is modulated due to spontaneous symmetry breaking, with the higher singlet density indicated with 
green plaquettes in the illustration to the left. In the 3D system, we add a Heisenberg interlayer coupling $J_\perp$.}
\label{fig:fig1}
\end{figure}

\subsection{Three-dimensional effects}

Assuming that the interactions in SrCu$_2$(BO$_3$)$_2$ are not significantly anisotropic in spin space (and there are no indications to the contrary as far as 
we are aware), the finite-temperature AFM ordering should be induced by weak interlayer couplings (given that a 2D with isotropic Heisenberg interactions
orders only at $T=0$). The ordering temperature $T_{\rm N}$ depends logarithmically on the interlayer coupling $J_\perp$; 
$T_{\rm N} \propto J \ln^{-1}(J_{\rm 2D}/J_\perp)$ \cite{Chakravarty1989,Irkhin1998}, where $J_{\rm 2D}$ should be interpreted as an effective 2D 
energy scale in a system with more than one intralayer coupling constant. Thus, the transition temperature can be a substantial fraction of $J_{\rm 2D}$ even for a 
very weak interlayer coupling, as confirmed explicitly by QMC calculations \cite{Sengupta2003}. In the context of a possible DQCP transition in 
SrCu$_2$(BO$_3$)$_2$, a crucial question is then how the interlayer couplings will affect the $T=0$ transition, and how this transition evolves 
to a finite-temperature transition.

Given that the 2D quantum phase transition between a $Z_2$ PS and an O(3) AFM is most likely weakly first-order, with DQCP characteristics up to some 
length scale, and that the CBJQ model exhibits this kind of behavior with emergent O($4$) symmetry, it would be interesting to focus experimental 
studies of SrCu$_2$(BO$_3$)$_2$ on detecting this symmetry. A critical question is then to what extent the emergent symmetry survives in the presence 
of the 3D couplings---the exact magnitudes of which are not known (though $J_\perp/J_{2D}$ of order $10^{-2}$ was estimated in Ref.~\cite{JingGuo2020}).

As mentioned above, reliably calculating ground state and $T>0$ properties is already
very challenging for the 2D SS model, and including the interlayer couplings is clearly beyond the scope of current DMRG and tensor-network methods
\cite{Lee2019,TNxiang} (though we note that a self-consistent mean-field could in principle be used to approximate these couplings, as a generalization
of the chain and multi-chain mean-field approaches \cite{Sandvik1999}). In this paper we instead follow up on the previous work on universal
aspects of the PS--AFM transition within the 2D CBJQ model, studying a 3D version of this model in the regime of very weakly coupled planes. The model and 
its quantum phases are illustrated in Fig.~\ref{fig:fig1}. We will extract the quantitative phase diagram using QMC simulations and investigate order parameter 
distributions as the PS--AFM transition is crossed at different temperatures and at different values of an interplane Heisenberg coupling. While we do 
not address any specific experiment, our observations allow us to judge whether a near-O($4$) symmetry is sufficiently established to have experimental 
consequences. Our main conclusion is that, while the emergent symmetry eventually is violated when the interlayer coupling is turned on, there should still
be detectable remnants of O(4) symmetry in the coexisting order-parameter fluctuations---on length scales up to tens or hundreds of lattice spacings---for
couplings of the magnitude expected in SrCu$_2$(BO$_3$)$_2$. We therefore expect that experiments such as inelastic neutron scattering, Raman scattering and thermodynamic measurements should be able to detect a corresponding low-energy mode.

\subsection{Paper outline}

We will compute the phase diagram of the 3D CBJQ model in the parameter space of temperature $T$, and intra-plane coupling ratio, and interplane coupling.
To extract the phase boundaries, we analyze the heat capacity as well as Binder cumulants defined with the PS and AFM order parameters. We focus on the regime
where the PS and AFM phase boundaries approach each other at weak interlayer coupling. Here we find that the phase transitions become first-order; thus
the PS, AFM, and paramagnetic phases come together at a triple point. We then study the symmetry properties of the joint PS and AFM order parameter
distribution close to the triple point.

The rest of the paper is organized as follows: In Sec.~\ref{sec:ii} the 3D CBJQ model and the QMC computed observables are defined. In Sec.~\ref{sec:iii} 
the phase diagram is extracted using finite-size scaling methods. Sec.~\ref{sec:iv} presents order parameter histograms, which reveal the O(4) symmetry aspects
of the first-order transitions at weak interlayer couplings, and quantitative measures of the degree of O(4) violation extracted from the same. In Sec.~\ref{sec:v}
we summarize our findings and discuss further implications and prospects for experimental studies of the PS--AF transition in SrCu$_2$(BO$_3$)$_2$
and future directions.

\section{Model and Method}
\label{sec:ii}

\subsection{3D CBJQ model}
\label{sec:ideas}
As shown in Fig.~\ref{fig:fig1}, we consider a 3D CBJQ model with interlayer coupling $J_{\perp}$ connecting 2D CBJQ layers with the following Hamiltonian
\begin{equation}
H = - J\sum_{\langle i j \rangle}P_{ij}-Q \hskip-2mm \sum_{ijkl \in \Box^\prime}  \hskip-1mm (P_{ij} P_{kl}+P_{ik} P_{jl})-J_\perp\sum_{\langle ij \rangle_{\perp}}P_{ij},
\label{Eq:CBJQModel}
\end{equation}
where $P_{ij}=(1/4-{\bf S}_i \cdot {\bf S}_j)$ is the singlet projector on sites $i,j$. In the $J$ sum  $\langle ij \rangle$ stands for the
nearest-neighbor sites within the layers, and in the $J_\perp$ term $\langle ij \rangle_{\perp}$ stands for interlayer nearest neighbors. In the $Q$ 
term, $ijkl \in \Box^\prime$ denotes the corners of $2\times2$ plaquettes forming a checkerboard pattern within the layers, with the same set of 
plaquettes chosen for all the layers (the red plaquettes in the right part of Fig.~\ref{fig:fig1}). We study square $L\times L$ layers with even 
$L$ and set the number of layers to $L_z=L/2$ in order to reduce the computational effort. The aspect ratio $L_z/L<1$ can also be expected to be
favorable in finite-size scaling in a highly anisotropic system \cite{Sandvik1999}. The boundary conditions are periodic in all directions. We
will use $N=L\times L\times L/2$ for the total number of lattices sites.

The two symmetry-breaking phases of the 3D CBJQ model are illustrated in  Fig.~\ref{fig:fig1} for a fixed interlayer interaction $J_{\perp}$. 
We define an intraplane coupling ratio $g=J/(J+Q)$ that we use as tuning parameter. In the simulations we set $J+Q=1$ as the energy scale and study two 
different interplane couplings, $J_\perp=0.01$ and $0.1$. The simulations are carried out using the stochastic series expansion (SSE) QMC
method~\cite{Sandvik1999,Sandvik2010} with lattice sizes up to $L=72$ in steps of $\Delta_L=8$. 

\subsection{Observables and finite-size scaling}

To analyze the different phases and transition between them, several physical observables are implemented in the SSE-QMC simulation.
First, order parameters $m_z$ for AFM and $m_p$ for PS phase are defined as
\begin{eqnarray}
m_z & = &\frac{1}{N}\sum_{i}(-1)^{(x_i+y_i+z_i)}S^z(i), \label{eq:eq2} \\
m_p & = & \frac{2}{N}\sum_{q}\phi(q)\Pi^z(q), \label{eq:eq3}
\end{eqnarray}
where $(-1)^{(x_i+y_i+z_i)}$ in Eq.~\eqref{eq:eq2} is the staggered phase factor and in Eq.~\eqref{eq:eq3} $\Pi^z(q)$ is a quantity appropriate for 
detecting the plaquette modulation in the PS state,
\begin{equation}
\Pi^z(q)=S^z(q)S^z(q+\hat{x})S^z(q+\hat{y})S^z(q+\hat{y}+\hat{x}),
\label{eq:piqzdef}
\end{equation}
with $q$ labeling the $N/2$ different $2\times 2$ plaquettes containing the $Q$ couplings. The PS symmetry-breaking is captured
with the phase factor $\phi(q) = \pm 1$ alternating on even and odd rows of the layers. Other operators can also been used to detect
plaquette order, as discussed, e.g., in Refs.~\cite{BWZhao2018,Takahashi2020}. We have also tested other options
for the 3D CBJQ model, but all results presented in this work are based on Eq.~\eqref{eq:piqzdef}.

\begin{figure}[t]
\includegraphics[width=\columnwidth]{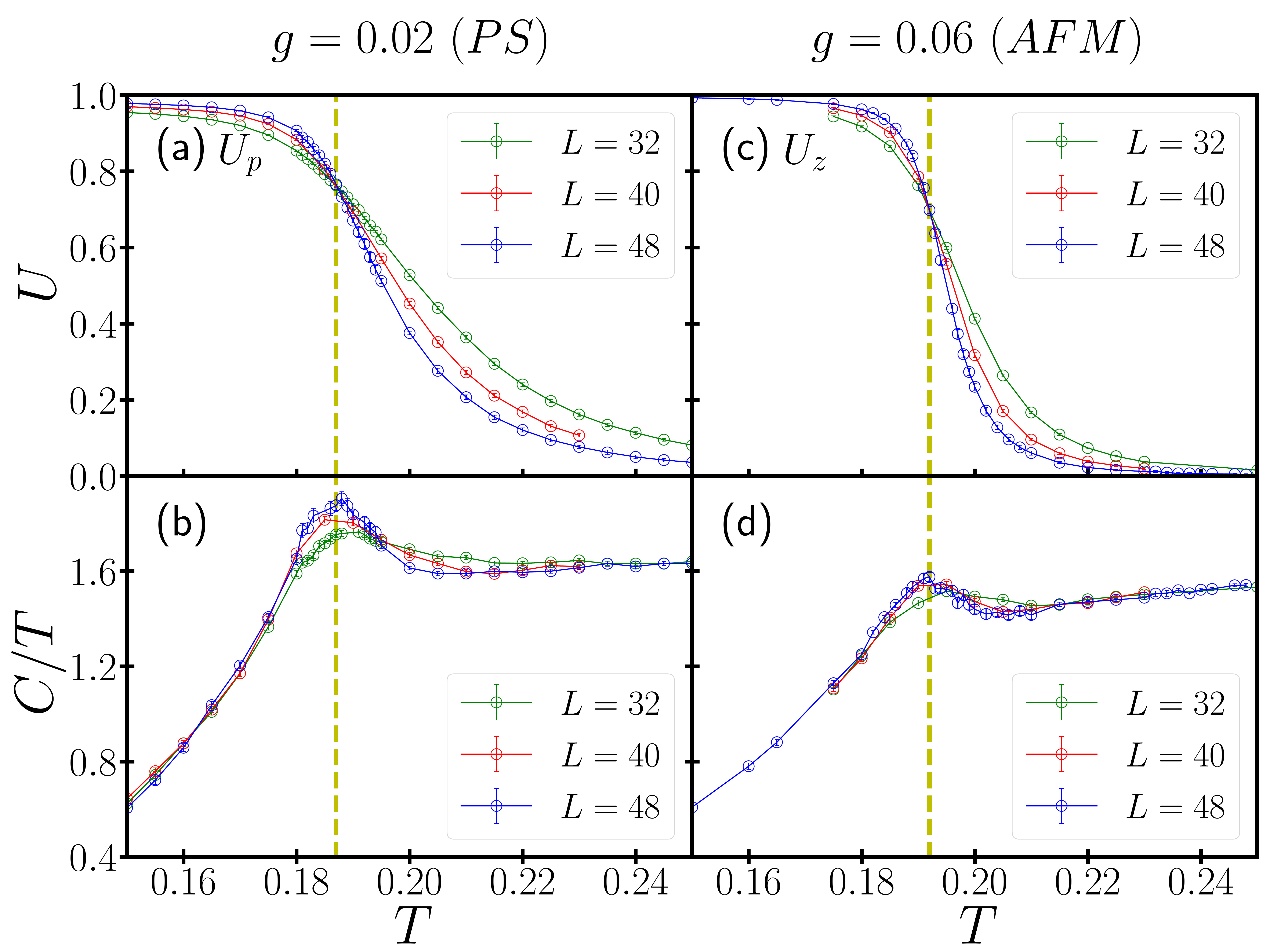}
\caption{Observables used to detect the temperature-driven  phase transitions into the PS phase at coupling ratio $g=0.02$ in panels (a), (b) and
into the AFM phase at $g=0.06$ in (c), (d). In both cases the inter-layer coupling of the 3D CBJQ model is $J_\perp = 0.1$ and results are shown for
lattice sizes $L= 32$, $40$, and $48$. The Binder cumulants corresponding to the respective order parameters are shown in (a) and (c), while the
heat capacity $C$ divided by $T$ is shown in (b) and (d). The  dashed lines mark the transition temperature determined from the crossing points of the
Binder ratios for different system sizes as consistently manifested in the location of the peak in $C/T$.}
\label{fig:fig2}
\end{figure}

Using the above order parameters, the corresponding Binder
cumulants are defined as
\begin{eqnarray}
U_z & = & \frac{5}{2}\left (1-\frac{\langle m_z^4\rangle}{3\langle m_z^2\rangle^2} \right ), \\
U_p & = & \frac{3}{2}\left (1-\frac{\langle m_p^4\rangle}{3\langle m_p^2\rangle^2} \right ),
\end{eqnarray}
where the coefficients are chosen to ensure  $U_z\to 1, U_p\to 0$  in the AFM phase and $U_z \to 0, U_p\to 1$ in the PS when $L\to \infty$.  

\begin{figure*}[t]
\includegraphics[width=\textwidth]{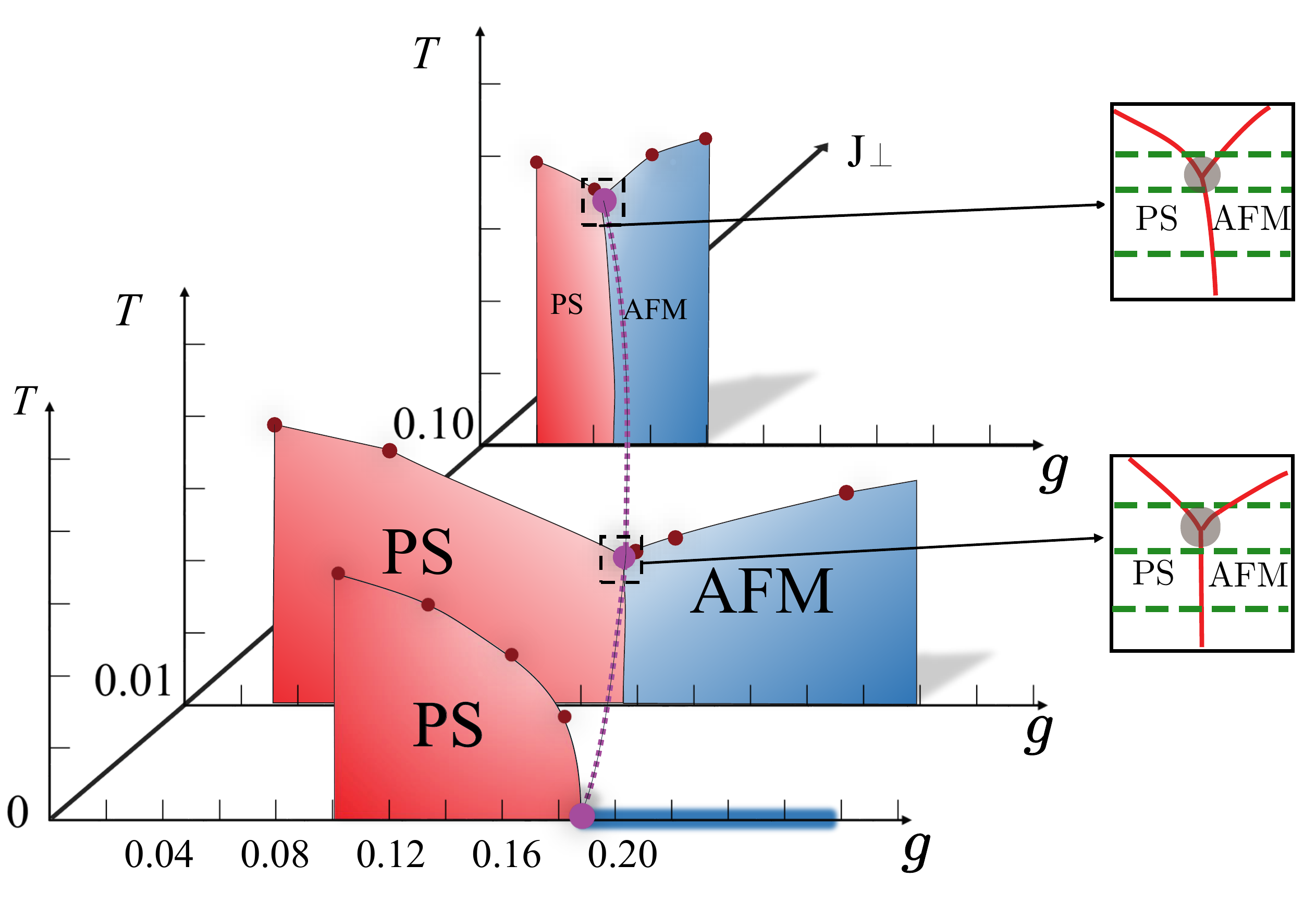}
\caption{ Phase diagram of the 3D CBJQ model based on SSE-QMC simulations and finite-size scaling analysis at $J_\perp =0$, $0.01$, and $0.1$. When  $J_\perp =0$, 
the long-range AFM state only exists at zero temperature, while for $J_\perp =0.01$ and $0.1$ it extends to finite  temperature. The small brown solid dots show 
transition points obtained from Binder cumulants crossing points in scans vs temperature, as exemplified in Fig.~\ref{fig:fig2}.  The position of larger purple solid 
dot at $T=0$ in the $J_{\perp}=0$ plane show the location of the PS--AFM quantum phase transition of the 2D CBJQ model determined in previous work~\cite{BWZhao2018}.
In the $J_{\perp}=0.01$ and $0.1$ panels, the triple points (purple dots) at which all three phases meet at $T>0$ (and which evolve from the quantum critical point at
$J_\perp=0$) are obtained from scans vs $g$ at fixed $T$, as  illustrated in the framed magnifications of the relevant temperature regions. The green dashed lines 
correspond to $T=0.06$, $0.085$, and $0.09$ for $J_\perp=0.01$ and $T=0.1$, $0.162$, and $0.168$ for $J_\perp=0.1$. Based on results such as those in 
Figs.~\ref{fig:fig4} and ~\ref{fig:fig5}, the triple-point is located between the two upper green dashed lines, for $J_{\perp}=0.1$ at $T=0.165(2)$ and 
for $J_{\perp}=0.01$ at $T=0.088(2)$ (and the corresponding $g$ values are $g \approx 0.041$ and $0.16$, respectively).} 
\label{fig:fig3}
\end{figure*}

With these observables, we can determine the finite temperature phase boundary between the PS and AFM phases at low temperatures as well as
the boundary between these two ordered phases and the paramagnet at higher temperature. Examples are illustrated in Fig.~\ref{fig:fig2}, where
we monitor the $T$ dependence of $U_p$ for $g=0.02$ in (a) and  that of $U_z$ for $g=0.06$ in (b), with the interlayer coupling fixed at $J_\perp=0.1$
in both cases. The almost size independent crossing points of the Binder cumulants clearly reveal the critical temperatures $T_c$. By scanning this way
at several different $g$ values we construct the phase boundaries. The so determined phase diagram is shown in Fig.~\ref{fig:fig3} and will be further
discussed in Sec.~\ref{sec:iii}.

An important observable for experimental studies of finite-temperature phase transition is the heat capacity, 
\begin{equation}
C=\frac{d\langle H\rangle}{dT},
\label{specific}
\end{equation}
which at a critical point in the thermodynamic limit scales as
\begin{equation}
C(T)\sim \left |1-\frac{T}{T_c}\right |^{-\alpha},
\label{cscaling}
\end{equation}
which translates to finite-size scaling at the critical temperature of the form
\begin{equation}
C(T_c,L)\sim L^{\alpha/\nu},
\label{cscaling}
\end{equation}
where $\nu$ is the exponent governing the divergence of the correlation length. In SSE-QMC simulations, $C$ can be obtained from the variance of
the expansion order $n$ as ~\cite{Sandvik1999Hubbard},
\begin{equation}
C= \frac{1}{N}(\langle n^{2}\rangle-\langle n\rangle^{2}-\langle n\rangle),
\label{specific}
\end{equation}
though in practice the statistical fluctuations are large and long runs are required to obtain useful results, especially at low temperatures.

We note that the recent mapping of the quantum phase boundaries SrCu$_2$(BO$_3$)$_2$ were based on the small heat capacity peaks at temperatures below
a broad maximum \cite{JingGuo2020}, and subsequently it was demonstrated that the broad maximum evolves into a critical point analogous to the
gas-liquid critical point at temperatures above the ordered phases \cite{Jimenez2020}. Here we will mainly use the Binder cumulants to extract the phase
boundaries, but we also demonstrate that our methods in principle can distinguish the different universality classes of the Ising-like PS transition
and the O(3) AFM transition.

Provided that the transitions are indeed continuous, for $J_\perp >0$ the transition from the paramagnet to the PS phase should be in the 3D Ising
universality class, while that out of the AFM phase should be in the 3D Heisenberg universality class. In the former, the critical exponent $\alpha$
is positive,  $\alpha=0.11$~\cite{3Dising}, and the heat capacity diverges at $T_{c}$ (here as a function of the system size). 
In the latter case, we expect only a cusp singularity and
no divergence on account of the negative exponent, $\alpha=-0.12$~\cite{3Dheisenberg}. The results shown in Fig.~\ref{fig:fig2}(c) and \ref{fig:fig2}(d)
are consistent with the expectations, with a weakly size-dependent peak height at the PS transition in Fig.~\ref{fig:fig2}(c) and essentially
size-converged peak at the AFM transition in Fig.~\ref{fig:fig2}(d). The peak locations coincide with the critical temperatures determined from
the Binder cumulants.

\subsection{Order parameter distribution}
\label{sec:iic}

As shown in Fig.~\ref{fig:fig3}, the phase transitions of the PS or AFM states to the high-temperature paramagnet meet at a point $(g,T)$ that depends
on the inter-layer coupling. The expectation from studies within the framework of classical O(N) models~\cite{Aharony2002,Aharony2002OldAN}, which would apply here for $T>0$, is that
there should be no emergent higher [O(4)] symmetry at this meeting point between $Z_2$ and O(3) symmetry-breaking transitions even if both transitions
remain continuous up to the (in that case) multi-critical point. As discussed in Sec.~\ref{sec:intro_emergent}, at $J_\perp=0$, the previous study of the
$T=0$ PS--AFM transition of the 2D CBJQ model \cite{BWZhao2018} revealed a first-order transition with a surprising emergent symmetry--the transition is
akin to the spin flop transition in a system with O(4) symmetry that is perturbed by an Ising-like anisotropy. However, there is no microscopic O(4)
symmetry in the CBJQ model, and the cause of the emergent symmetry is presently unclear. In the simplest scenario, the first-order transition is close to
a quantum-critical point with O(4) symmetry \cite{Serna2019}, and the length-scale on which this symmetry survives approximately away from the critical point is
very large (at least up to 100 lattice spacings in the case considered). A very large length scale of emergent supersymmetry has also been demonstrated
in the vicinity of certain topological phase transitions in fermionic systems  \cite{Yu2019}.

Though we do not expect any asymptotic emergent symmetry in the 3D CBJQ model, it is interesting and experimentally relevant to investigate how the O(4)
symmetry---whether asymptotically present or persisting only up to some very large length scale in the 2D limit---evolves as the direct AF--PS transition
moves up to $T>0$ as $J_\perp$ is turned on and eventually terminates at the multi-critical or triple point. Thus, we will examine symmetry relationships
between the PS and AFM order parameters. To this end, in the simulations we accumulate the joint probability distribution $P(m_z,m_p)$ of the order parameters
$m_z$ and $m_p$ defined above in Eqs.~(\ref{eq:eq2}) and (\ref{eq:eq3}), with point pairs $(m_z,m_p)$ obtained in equal-time measurements on the SSE 
configurations. Note that $m_z$ and $m_p$ are both diagonal in the basis used in the simulations. In Sec.~\ref{sec:iv} we will further discuss quantitative
measures of O(4) symmetry that we use to examine the fate of the higher symmetry on increasing the length scale (the system size).

\section{Phase diagram and triple point}
\label{sec:iii}

With the observables and analysis outlined in Sec.~\ref{sec:ii}, we obtain the complete phase diagram of 3D CBJQ model spanned by the axes
of $g$-$T$-$J_{\perp}$ in Fig.~\ref{fig:fig3}. While we have only studied two values of the inter-layer coupling, $J_\perp=0.1$ and $0.01$,
the overall qualitative behavior is still clear. For a  fixed $J_\perp$, along the $g$ direction the phase diagrams in Fig.~\ref{fig:fig3} comprise
the low-temperature PS (the red area) and AFM (the blue area) phases, as well as the featureless paramagnetic phase at higher temperatures. We do not
anticipate any other phases as $J_\perp$ is increased to much larger values, but the PS phase will eventually vanish.
We here focus on small $J_\perp$, as expected in the SS material.

In the 2D case, $J_\perp = 0$, the long-range AFM phase only exists exactly at $T=0$, with the spin correlation length diverging exponentially as
$T \to 0$ in the well known ``renormalized classical'' region of the 2D AFM Heisenberg systems \cite{Chakravarty1989}. Even for very small $J_\perp > 0$,
the AFM transition temperature $T_{\rm N}$ is substantial, however, on account on the logarithmic form related to the exponential behavior of the correlation
length in the 2D layers \cite{Chakravarty1989,Irkhin1998,Sengupta2003}. The logarithmic form is also reflected in the rather small increase in $T_{\rm N}$
when increasing $J_\perp$ from $0.01$ to $0.1$ (once the system is inside the AFM phase), as seen in the phase diagram in Fig.~\ref{fig:fig3}.

Our main interest here is in the direct AFM--PS transition, especially close to the end point where the three phases meet. As indicated in
Fig.~\ref{fig:fig3}, and as we will elaborate on further below, all three phase boundaries appear to be first-order transitions in this regime;
hence the meeting point should be classified as a triple point. We expect such a triple point also in SrCu$_{2}$(BO$_{3}$)$_{2}$, but current experiments 
have not yet been able to explore it because of experimental limitations. According to
Refs.~\onlinecite{JingGuo2020} and \onlinecite{Jimenez2020}, the AF--PS transition should be located between 2.6 GPa and 3 GPa based on the phase boundaries obtained
in temperature scans of the heat capacity. These scans detected the phase boundaries between the AF or PS phases and the paramagnetic phase depending
on the pressure. However, between 2.4 and 3 GPa the transition temperatures are below the lowest accessible temperature, $T \approx 1.5$ K,
in the experiments in Ref.~\onlinecite{JingGuo2020}. In Ref.~\onlinecite{Jimenez2020} lower temperatures were reached but only up to 2.65 GPa.

There are disagreements in the literature on the nature of the PS--AF quantum phase transition in the 2D SS model. A clearly first-order transition
was obtained with tensor-product states in Ref.~\cite{sstensor}, while a continuous DQCP transition was argued based on DMRG results in Ref.~\cite{Lee2019}.
These methods are affected by finite tensor dimension and limited lattice sizes, respectively. Given the $Z_2$ symmetry of the PS state, a truly continuous
DQCP transition appears unlikely \cite{Block2013,Nakayama2016}, though the system could very well be extremely close to such a point, thus with such
a weak first-order transition that it appears to be continuous on the small lattices accessible with the 2D DMRG method. The first-order transition
with emergent O(4) symmetry in the 2D CBJQ model also may point to the proximity of a DQCP in this case. Given all these results, it seems likely that the
transition in the 2D SS model is weakly first order and hosts emergent O(4) symmetry up to large length scales. Given the inter-layer couplings present in the SS
material, an important question is then what remnants there are of emergent O(4) symmetry on the PS--AF transition as it extends up from $T=0$ and
ends at the triple point. Our study here is primarily aimed at answering this question in the context of the 3D CBJQ model, where large-scale QMC
calculations can be carried out, and we expect the results to be relevant to SrCu$_{2}$(BO$_{3}$)$_{2}$.

Studying the triple point is by no means an easy task, as it is very difficult to locate it in the plane $(g,T)$ at finite $J_\perp$ within reasonable computer
resources. The finite-size shifts of the point and the scaling corrections to the asymptotic flows of observables are significant; thus large system sizes are
required. To approximately locate the triple point, we first carried out temperature scans for different system sizes, as already illustrated above in
Fig.~\ref{fig:fig2}. In this way we obtain points on the PS--paramagnet and AF--paramagnet phase boundaries, and it becomes apparent roughly where the triple
point is located for given $J_\perp$. We then also carried out scans versus
$g$ for fixed $T$ in the close neighborhood of the estimate temperature of the triple point, as illustrated in the insets of Fig.~\ref{fig:fig3} and 
explained below.

Most of the computational resources were devoted to the $g$ scans, 
where our goal is to locate points on the first-order line close to the triple point
and to detect the splitting of the line into two separate transitions as $T$ exceeds the temperature of the triple point. We focus on the Binder cumulants
$U_\alpha$ ($\alpha=z,p$), interpolated in our data set for different system sizes to extract crossing points $g_{\alpha,c}$ at which
$U_{\alpha}(g_{\alpha,c}(L),L)=U_{\alpha}(g_{\alpha,c}(L),L+8)$  (the method first discussed by Luck \cite{luck1985corrections}; recent tests and
further discussion can be found in Ref.~\cite{Shao2016}).  We extract such size-dependent transition points
$g_{\alpha,c}(L)$ for both AF ($\alpha=z$) and PS $(\alpha=p)$ order. As long as the temperature is at or below the triple point, the two transition points
should extrapolate to the same value $g_{z,c}(L) \to g_{p,c}(L)$ for $L \to \infty$. In principle, for a continuous transition the convergence would
be of the form $g_{c\alpha}(L) - g_{c\alpha}(\infty) \sim aL^{-b}$, where $b={1/\nu+\omega}$ ($\omega$ being the leading correction exponent)
while for the first-order transition expected here we should have $b=D$, the spatial dimensionality (when $T>0$). However, it is well known that the exponents
in practice often are ``effective'' ($L$-dependent) for the system sizes that can be reached. We will not focus on the values of the exponent, but just carry out
power-law fits to $g_{\alpha,c}(L)$ for the largest system sizes available. Using a few of the largest system sizes (in pairs $L$ and $L+8$) we often find
almost linear behaviors in $1/L$, and we expect that the curves will further flatten out with increasing system size. Our main focus is on whether the data
based on AF and PS orders approach each other or deviate further as $L$ increases, as a signal of a direct transition or two separate transitions.

\begin{figure}[htp!]	
\includegraphics[width=\columnwidth]{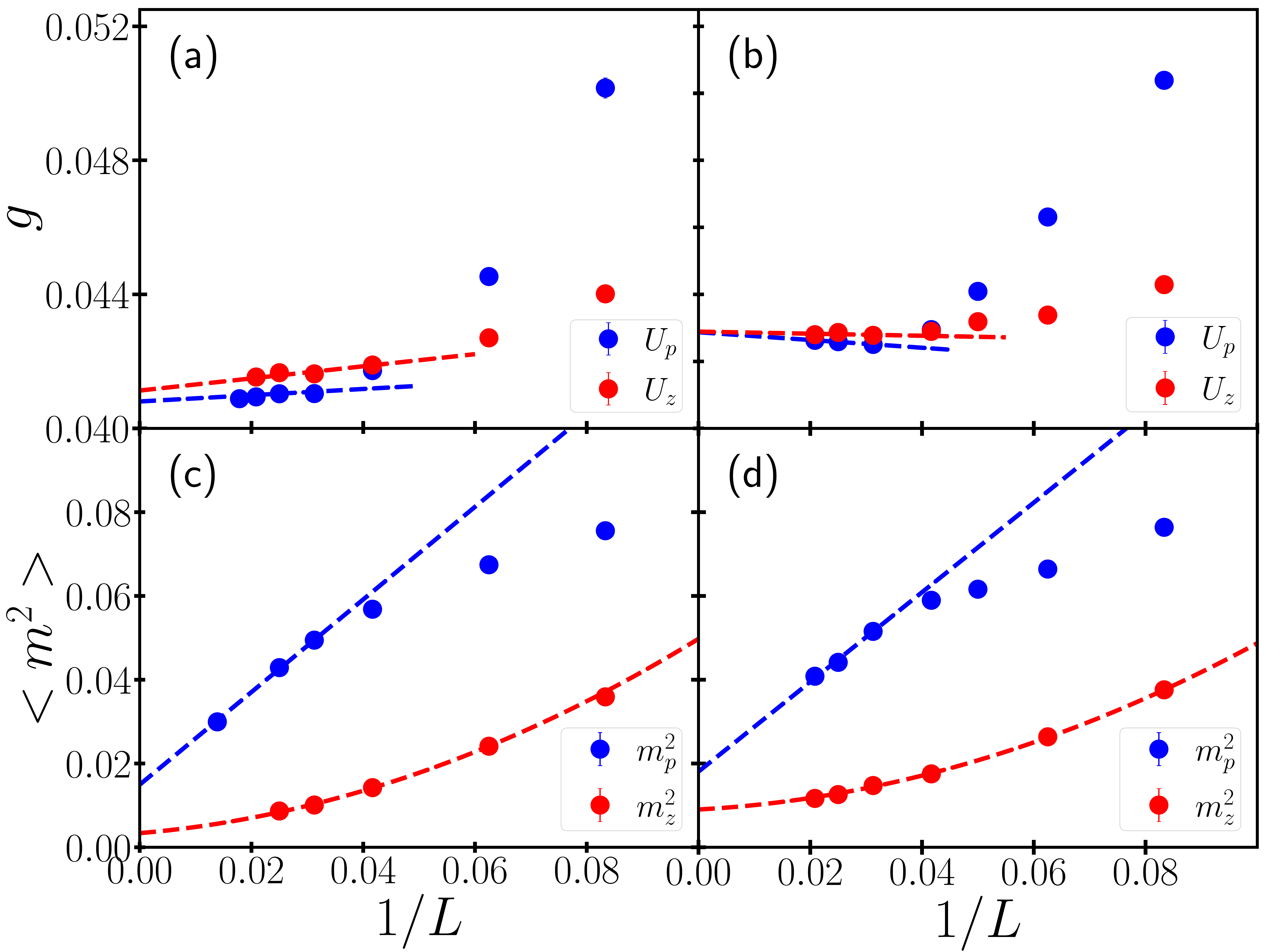}
\caption{Finite-size scaling results in the close neighborhood of the triple point for $J_\perp = 0.1$. The temperature is $T=0.168$ in the left panels and
$T=0.162$ in the right panels. (a) and (b) show transition points extracted from Binder cumulants crossings for system sizes $L$ and $L+8$, graphed as circles
vs $1/L$. The dashed lines show fits to the data for the four largest $L$ values. Panels (c) and (d) show the two squared order parameters of the size $L+8$
system extracted at those same crossing values of $g$, along with polynomial fits. The non-zero extrapolated order parameters imply first-order phase
transitions.}
\label{fig:fig4}
\end{figure}

\begin{figure}[htp!]
\centering
\includegraphics[width=\columnwidth]{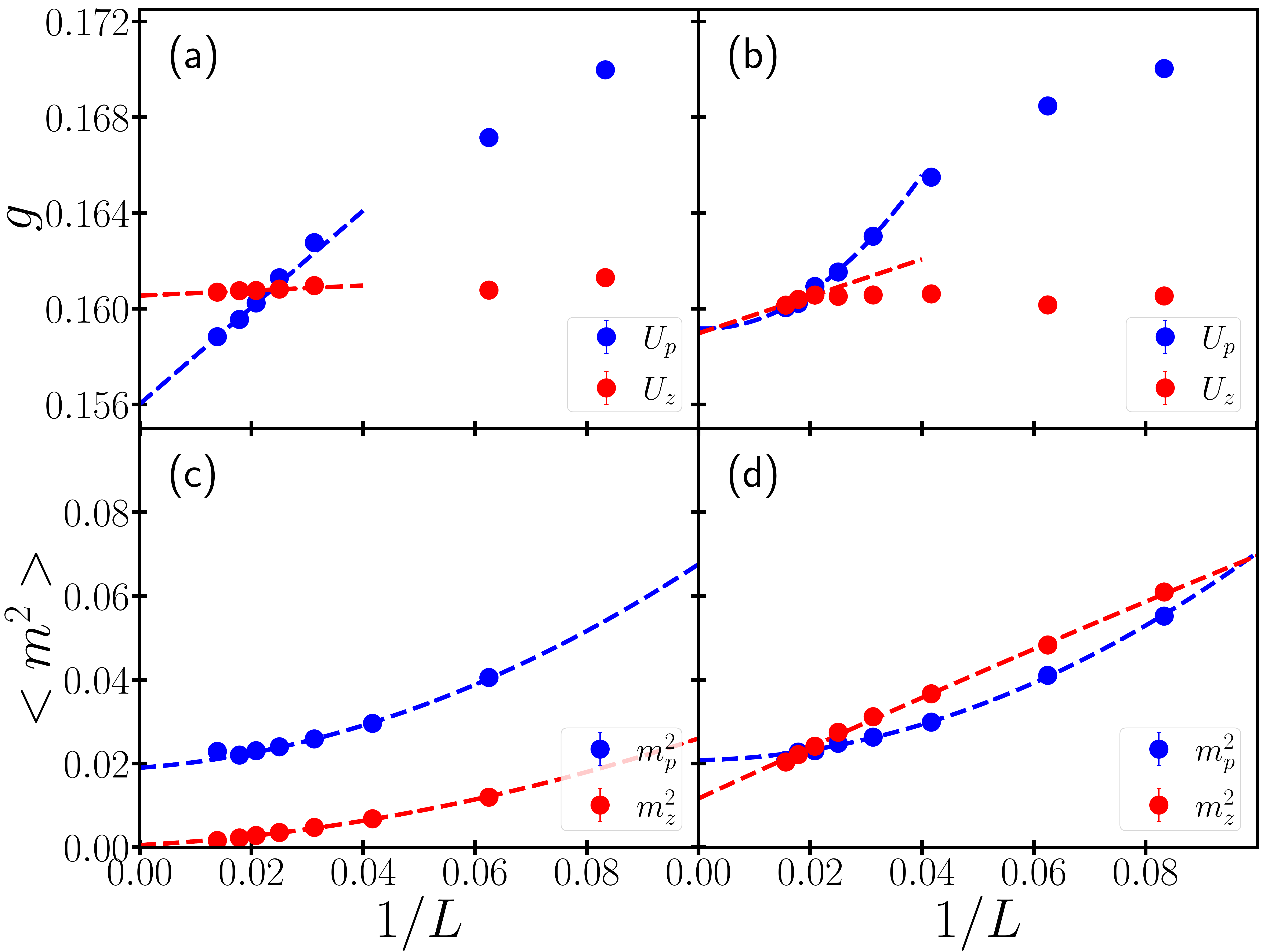}
\caption{Results analogous to those in Fig.~\ref{fig:fig4} but at a much lower inter-layer coupling; $J_\perp=0.01$. The temperature is $T=0.09$ and
$0.085$ in the left [(a) and (c)], and right [(b) and (d)] panels, respectively. Also in this case, the order parameters extrapolate to non-zero values,
indicating first-order transitions.}
\label{fig:fig5}
\end{figure}

Figures \ref{fig:fig4} and \ref{fig:fig5} show illustrative results for $J_\perp=0.1$ and $0.01$, respectively. In each of these figures we show data for two
different choices of the temperature; one [Figs.~\ref{fig:fig4}(a) and \ref{fig:fig5}(a)] where the PS and AF transition points separate clearly from each other
for increasing $L$, thus indicating two separate transitions, and one [Figs.~\ref{fig:fig4}(b) and \ref{fig:fig5}(b)] where the points approach each other
and extrapolate to a common transition point to within the precision of our procedures. Based on these results, we estimate that the triple point is at 
$T_c = 0.165(2)$ ($g_c \approx 0.041$) when $J_\perp =0.1$ and $T_c=  0.088(2)$ ($g_c \approx 0.16$) when $J_\perp =0.01$.

Next we investigate the nature of the transitions close to the triple point. At a continuous phase transition, the order parameter(s) approach  zero at
the critical point in the thermodynamic limit, while at a first-order transition coexistence of two (or three at a triple point) phases implies non-zero values
of the order parameters concerned. In our study we interpolate the squared order parameter $m_{p}^{2}(g^*)$ [$m_{z}^{2}(g^*)$] at the cumulants crossing points 
$g=g^*$ where $U_{p}(g_{cp}(L),L)=U_{p}(g_{cp}(L),L+8)$ [$U_{z}(g_{cz}(L),L)=U_{z}(g_{cz}(L),L+8)$], for the larger of the two system sizes; $L+8$. Then we extrapolate 
these order parameters by polynomials to the thermodynamic limit. The results at $J_{\perp}=0.1$ in Figs.~\ref{fig:fig4}(c) and \ref{fig:fig4}(d) show that the
PS and AF order parameters remain finite at their respective transition points at both temperatures, $T=0.162$ and $T=0.168$, when $L\to \infty$, implying
that all these phase transitions are first-order. Thus, indeed, the point of coexistence of the three phases is a triple point, not a multicritical point.
Similar convergence behaviors are also observed at $J_{\perp}=0.01$ in Figs.~\ref{fig:fig5}(c) and \ref{fig:fig5}(d) at $T=0.09$ and $T=0.085$, demonstrating
first-order transitions and a triple point also in these cases. At $T=0.09$, the extrapolated $\langle  m^2_z\rangle$ is small, suggesting that the transition
here is very weakly first order, close to the point where the AF--paramagnetic transition becomes continuous. We find behaviors consistent with continuous
$T>0$ PS and AF transitions in all cases when $g$ is sufficiently far from the triple point. On the direct PS--AF transition, we find more strongly first-order
behavior as $T$ decreases from the value at the triple point, as we will discuss further in Sec.~\ref{sec:iv}.

While we only considered a small part of the entire parameter space of the 3D CBJQ model, the phase diagram in Fig.~\ref{fig:fig3} contains the main
salient features of the model and we do not expect any other phases as long as the signs of all coupling constants are positive in Eq.~(\ref{Eq:CBJQModel}),
i.e., where sign-problem free QMC simulations are possible. When increasing $J_\perp$ further from the small values considered here, we expect the PS phase to
eventually vanish, on account of the clear shrinking of the phase when increasing $J_\perp$ from $0$ to $0.1$, and also because of the general expectation
when the interlayer coupling is the conventional Heisenberg exchange.

\section{Emergent symmetry}
\label{sec:iv}

As already discussed in Sec.~\ref{sec:intro_emergent}, in the 2D CBJQ model the $T=0$ quantum phase transition between the PS and AF states
is unusual, being similar to a spin-flop transition in an O(4) model with Ising type anisotropy. The transition is of first-order in the sense of a discontinuous
jump from the PS phase into the AFM phase, as in a conventional first-order transition, but there are no detectable tunneling barriers between the two phases.
In the coexistence state, the system can rotate its long-range order parameter without energy cost between the two phases.

In a classical O(4) model with magnetization components $(v_0,v_1,v_2,v_3)$, a deformation inducing $Z_2$ and O(3) symmetry-breaking phases [with order parameters
$v_0$ and $(v_1,v_2,v_3)$, respectively], there are similarly no free-energy barriers separating the trivially coexisting ordered states below $T_c$ at
the special O(4) point---the different states just correspond to different directions of the four-dimensional vector order parameter. Lacking a microscopic
O(4) symmetry, this kind of behavior is highly non-trivial in the CBJQ model, however, as the two order parameters do not correspond in any simple way to
different components of a four-dimensional vector. The higher symmetry must therefore be an emergent low-energy property.

Emergent symmetries at DQCP transitions and similar quantum-critical points have been actively studied recently
\cite{senthilfisher,Nahum2015b,Suwa2016,Wang2017,Gazit2018,Sreejith2019,HongYao2019,Sato2020}
and it is certainly possible that the observed emergent O(4) 
symmetry is simply indicating the proximity of the CBJQ model to a critical point with O(4) symmetry \cite{Serna2019}. The symmetry should then be broken on some
length scale larger than the largest lattice sizes, $L \approx 100$, that were studied. Such a large length scale would then still be surprising, given that the
transition is very clearly first-order and outside the realm of ongoing discussions of very weak DQCP-like first-order transitions versus truly asymptotically
continuous DQCP transitions \cite{Gorbenko2018,RMa2020,Nahum2020,SandvikCPL2020,Zhao2020}. The possibility remains that the coexistence state at the PS--AF
transition in the CBJQ model, as well as in a related 3D classical loop model \cite{Serna2019}, remains O(4) symmetric in the thermodynamic limit (though such a
scenario is outside current understanding of phase transitions). We note that a different 2D J-Q model with $Z_4$-breaking PS order exhibits emergent
O(5) symmetry \cite{Takahashi2020}.

In this work, we do not attempt to address the issues of continuous versus weakly first-order DQCP transitions or whether observed emergent asymmetries are
asymptotically present or not, but take a more practical approach of establishing the 3D CBJQ model as an example of a system with a PS--AF transition where
the violation of an emergent O(4) symmetry can be tuned from zero or extremely weak in the 2D limit to strong when the 3D coupling become significant. Reliable
bench-mark calculations for such a model will be useful, in particular, for planning and interpreting future experiment on the SS material SrCu$_{2}$(BO$_{3}$)$_{2}$.
We note that there is not a priori reason to expect an asymptotic emergent O(4) symmetry in the 3D CBJQ model, as the exotic physics related to the DQCP should
be particular to 2D systems. Nevertheless, with small interlayer couplings $J_\perp$ there should at least be some remnants of the 2D $T=0$ emergent symmetry on
the $T\ge 0$ PS--AF phase boundary, in  models as well as in SrCu$_{2}$(BO$_{3}$)$_{2}$, provided that the interlayer couplings are sufficiently weak in the
latter. We here characterize the the stability of the approximate O(4) symmetry in the 3D CBJQ model.

\subsection{Joint order parameter distribution}
\label{sub:histo}

As discussed in Sec.~\ref{sec:iic}, with the AF and PS order parameters defined in Eqs.~(\ref{eq:eq2}) and (\ref{eq:eq3}), respectively, we accumulate the
distribution $P(m_z,m_p)$. By the microscopic symmetries of the order parameters, we can completely characterize this distribution with the absolute values $|m_z|$
and $|m_p|$. We here show some examples of distributions close to the triple points determined in Sec.~\ref{sec:iii} for $J_\perp=0.1$ and $0.01$. Quantitative
analysis of the distributions will be presented below in Sec.~\ref{sub:omega}.

\begin{figure}[t]
\includegraphics[width=\columnwidth]{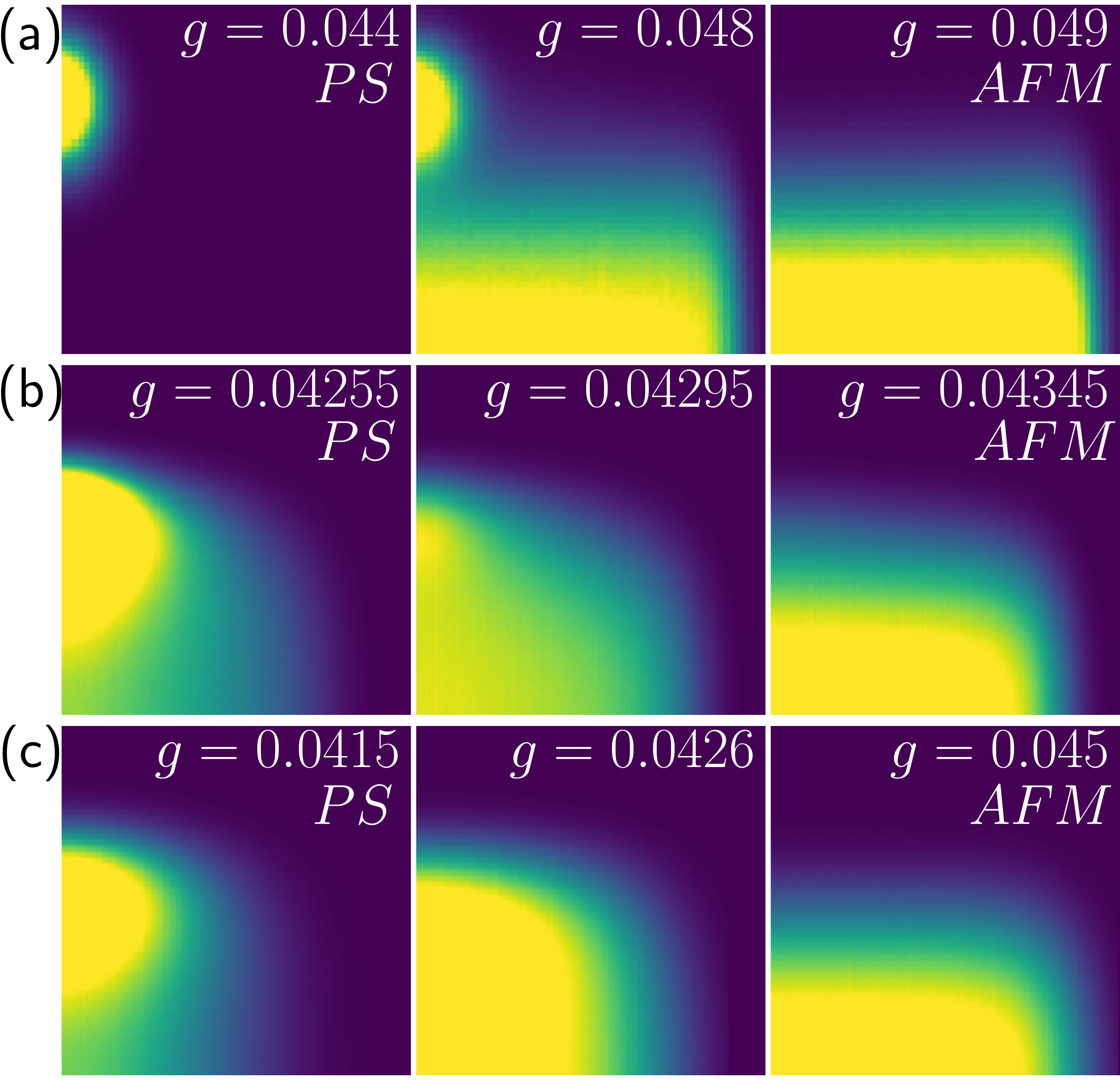}
\caption{Order parameter histograms $P(|m_p|,|m_z|)$ in the neighborhood of the PS--AF transition at $J_\perp=0.1$. For each of the three cases in (a), (b), and (c),
the temperature is held fixed and the $g$ values in the left and right panels correspond to the PS and AF phases respectively. The middle panel is for the g-value
where the coexistence features are seen most clearly. The system sizes and temperatures are $L=32$, $T=0.1$ (significantly below the triple point) in (a),
$L=56$, $T=0.162$ (slightly below the triple point) in (b), and $L=48$, $T=0.168$ (slightly above the triple point) in (c). The axis have been scaled to show
both order parameters roughly at equal magnitude in the middle panels (with the same scales using also in the left and right panels), and the color coding of the 
probability density is chosen on a linear scale adapted so that almost the full total weight is revealed clearly above the dark blue (zero) background.}
\label{fig:fig7}
\end{figure}

\begin{figure}[t]
\centering
\includegraphics[width=\columnwidth]{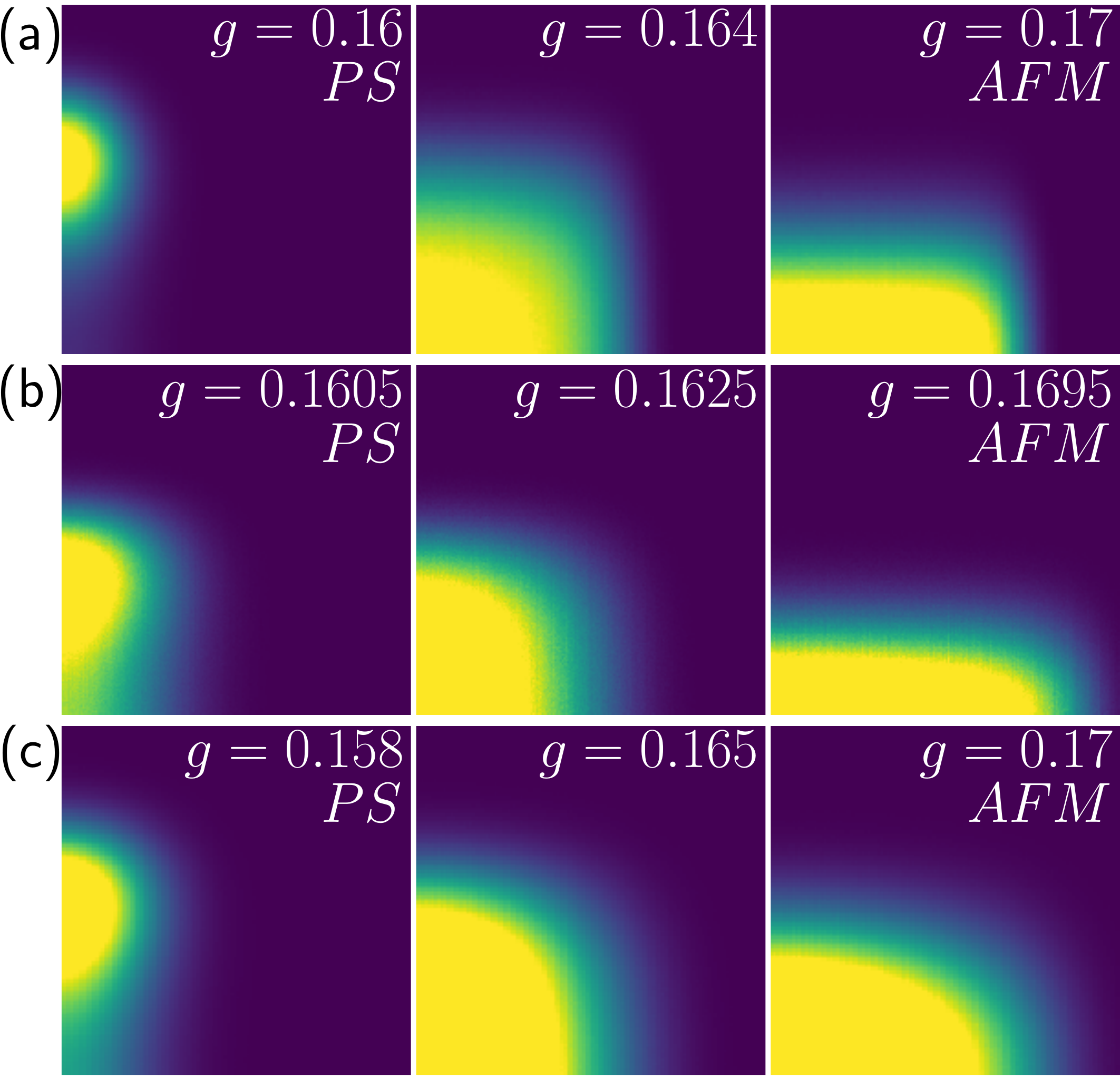}
\caption{Order parameter histograms $P(|m_p|,|m_z|)$ at $J_\perp=0.01$ graphed as in Fig.~\ref{fig:fig7} to show the evolution as the PS--AF transition is traversed.
In (a) $L=48$, $T=0.06$ (significantly below the triple point), in (b) $L=72$, $T=0.085$ (very close to but slightly below the triple point)
and in (c) $L=48$, $T=0.09$ (slightly above the triple point).}
\label{fig:fig8}
\end{figure}

At a conventional first-order phase transition, the joint distribution of two order parameters, e.g., $P(m_z,m_p)$ in the present case, will essentially become a
superposition of the two distributions pertaining to the individual non-zero order parameters on either side of the transition point. The AFM order parameter $m_z$
that we use here is one component out of three of an $O(3)$ symmetric order parameter, and when projected down to the $m_z$ axis the distribution $P(|m_z|)$ will 
therefore be uniform from $0$ up to the magnitude $m=(\langle m_x^2+m_y^2+m_z^2\rangle)^{1/2}$ of the vector order parameter in the AF phase. Since $m_p$ is a scalar 
order parameter, its
distribution in the PS phase in the thermodynamic limit will be a $\delta$-function at the non-zero value of the mean order parameter. Both distributions will
be broadened by finite-size fluctuations. When traversing a first-order PS--AF transition, the joint distribution $P(|m_z|,|m_p|)$ for a finite system should
therefore evolve from a peak on the $|m_p|$ axis in the PS phase to a broadened line on the $|m_z|$ axis in the AFM phase. In the narrow transition region
separating the two phases the distribution should exhibit both features. There will be some small weight connecting the two features in the distribution, representing
the free-energy barriers between the two phases (or energy barriers to tunneling at $T=0$). As the system size increases, the two phases become increasingly
isolated and the peak and line features sharpen as the finite-size fluctuations of the non-zero order parameters diminish.

An example of the evolution of $P(|m_z|,|m_p|)$ when crossing the PS--AF transition is shown in Fig.~\ref{fig:fig7}(a) for the $J_\perp=0.1$ system at $T=0.1$,
where the transition is rather strongly first-order. The superposition aspect of the distribution in the coexistence state is very clear. In Fig.~\ref{fig:fig7}(b)
results are shown at $T=0.162$, i.e., close to but still below the triple point. Here we still see both PS and AF features in the coexistence region, but
much less clearly than at the lower temperature. The weight between the features is much more pronounce, signifying a collective state in which the two order
parameters coexist simultaneously, as opposed to fluctuating between two macroscopically distinct states. Finally, in In Fig.~\ref{fig:fig7}(c) the temperature
is slightly above the triple point according to our results in Sec.~\ref{sec:iii}, and instead of coexistence there is then a region with none of the orders. In that
region, the joint distribution will eventually, in the limit of infinite system size,  become a $\delta$-function at $|m_z|=0,|m_z|=0$, but the system studied
here still has significant fluctuations and a broad maximum on account of the close proximity to the triple point.

In Fig.~\ref{fig:fig8} similar histograms are shown at the weaker interlayer coupling, $J_\perp=0.01$. Here the superposition features are not present in
the coexistence state, and instead the distributions in the transition region feature a single broad plateau. In the 2D CBJQ model, it was shown that the
coexistence distribution is that expected if the order parameter combine into an O(4) vector \cite{BWZhao2018}, which leads to a circular distribution
with uniform density when the fluctuations of the vector of non-zero mean length (i.e., uniformly distributed points on a four-dimensional sphere) is
projected down to two dimensions. It should be noted that $m_z$ is one component of an O(3) vector, and this symmetry is not broken in the simulations.
Therefore, the O(4) symmetry can be properly studied with just the two components considered here and in Ref.~\cite{BWZhao2018}. However, to reveal the
symmetry, if present, we have to take into account  that $m_z$ and $m_p$ have arbitrary factors in their definitions, and therefore a rescaling of one of
the two components has to be performed. In Fig.~\ref{fig:fig8} the scales of the $m_z$ and $m_p$ have been chosen so that the two order parameters
approximately equal magnitude in the coexistence state. In sec.~\ref{sub:omega} we will discuss the rescaling in more detail and present a quantitative
measure of the emergent symmetry.

\begin{figure*}[t]
\includegraphics[width=14cm]{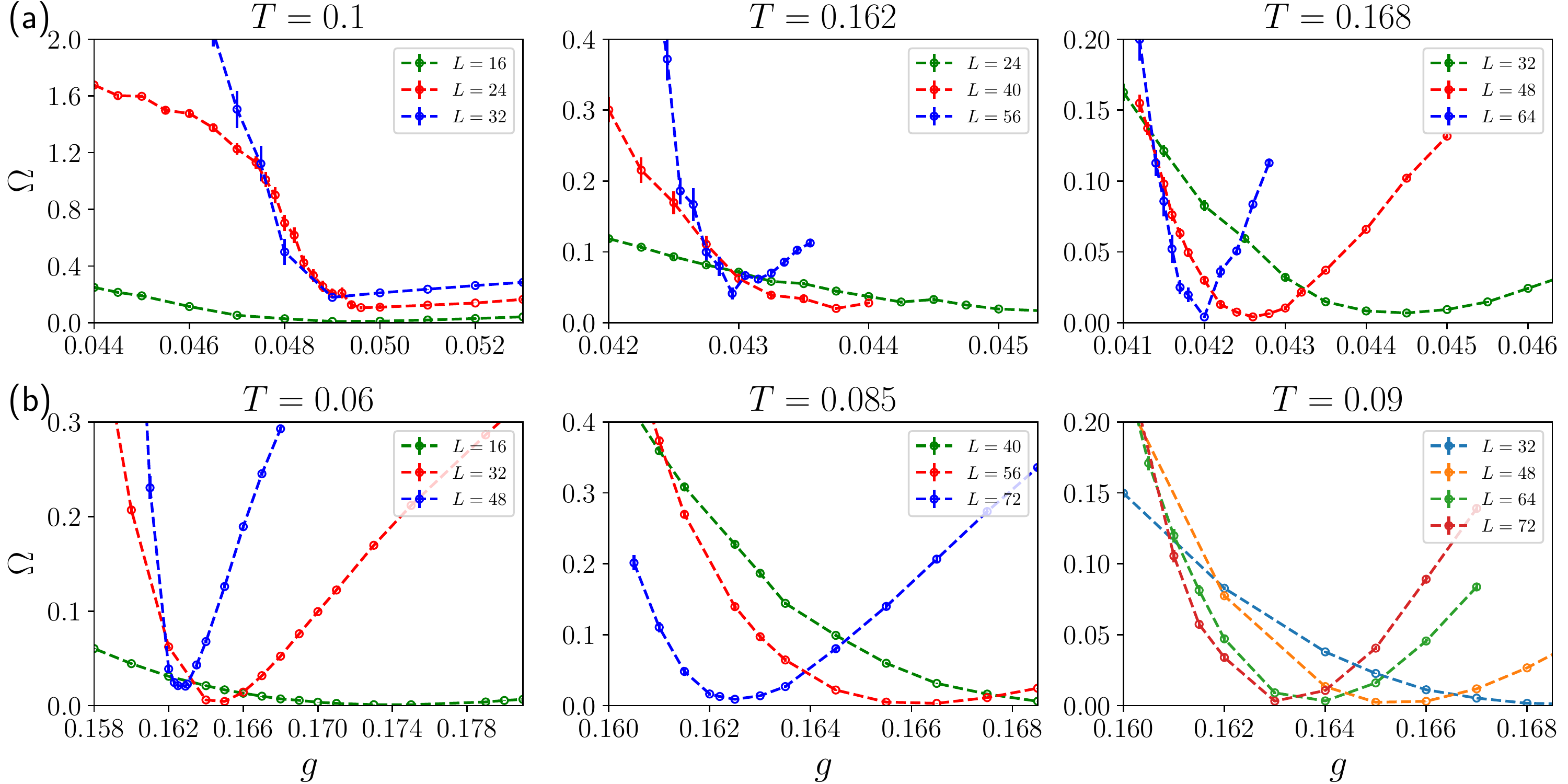}
\caption{The sum of frequency moments $\Omega$, defined in Eq.~(\ref{ffunction}),
which quantifies the deviation of the joint PS--AF order parameter distribution from perfectly
O(4) symmetric ($\Omega=0$). Results are shown versus $g$ for $J_\perp = 0.1$ in (a) and $J_\perp = 0.01$ in (b) at the same temperatures used in Figs.~\ref{fig:fig7}
and \ref{fig:fig8} as indicated on top of each graph. Multiple system sizes are used to illustrate the eventual violation of the O(4) symmetry on the first-order
lines below the triple point, seen as an increase in the minimum $\Omega$ value with $L$ at $T=0.1$ and $0.162$ in (a), and similarly in (b) at $T=0.06$ and
$0.085$. Above the triple point, $T=0.168$ in (a) and $T=0.09$ in (b), the minimum value stays very close to $0$ for all $L$ on the account of the paramagnetic 
state lacking long-range order.}
\label{fig:fig6}
\end{figure*}

The histograms in Fig.~\ref{fig:fig7} and Fig.~\ref{fig:fig8} give visual confirmation of what could be expected if the $J_\perp=0$ system has
emergent O(4) symmetry of the $T=0$ coexistence state: When $J_\perp > 0$ but small, the triple point is located at a low temperature and the emergent
symmetry survives up to some length scale on the entire $T\ge 0$ PS--AF transition line. At $J_\perp=0.1$, the system sizes accessible in our simulations
already exceed the cross-over length scale at which the coexistence state becomes conventional and the order parameter distribution shows distinct features originating
from both phases, which are separated by free-energy barriers. When $J_\perp = 0.01$, the coexistence state reflects an emergent near-O(4) symmetry up to the
largest system sizes studied. There are then no barriers, as the system can be continuously rotated between the phases at constant energy.

\subsection{Quantitative characterization of O(4) symmetry}
\label{sub:omega}

We here provide a quantitative analysis of the emergent O(4) symmetry and its break-down, using these integrals over the order parameter distribution
$P(m_z,m_p)$:
\begin{equation}
I_{2q} = \frac{1}{M} \sum_{k=1}^M \cos[2q\theta(\tilde m_z,\tilde m_p)_k],
\label{iqdef}
\end{equation}
where the point pairs $(\tilde m_z,\tilde m_p)_k$ are obtained from those sampled in the simulations, $(m_z,m_p)_k$, by a suitable rescaling
to put the two order parameters on equal footing so that a possible emergent O(4) symmetry can be revealed. After this rescaling the angle $\theta$
corresponding to the points is computed and used in Eq.~(\ref{iqdef}) for a large number (millions) of points.

We rescale one of the components by a factor $a$, $\tilde m_p = am_p$ (keeping $\tilde m_z = m_z$), with $a$ determined for a given system size and temperature at
the $g$ value where the cumulants for the two order parameters are equal, $U_p(L)=U_z(l)$ (which is a convenient single-size definition of the direct transition
point). The factor is fixed at this value for all values of $g$ for given $T$ and $J_\perp$. 
It is also possible to use a $g$-dependent rescaling factor $a(g)$, but in Ref.~\cite{BWZhao2018}
it was concluded that the minimums in the moments $|I_{2q}|$ in Eq.~(\ref{iqdef}), which signify the point closest to O(4) symmetry, are more pronounced if
the rescaling is fixed at the transition point. We refer to the Supplemental Information in Ref.~\cite{BWZhao2018} for a comprehensive discussion of the
rescaling.

We can study individual angular moments $I_{2q}$, as was done in the 2D, $T=0$ system in Ref.~\cite{BWZhao2018}, but for simplicity  we here 
form a measure combining the first four of them
\begin{equation}
\Omega= \sum_{q=1}^4 I_{2q}^2.
\label{ffunction}
\end{equation}
This quantity will detect deviations from the O(2) symmetry of the 2D distribution originating from any of the moments used, though in practice it is dominated by the
features in $I_2$ and $I_4$. These two moments are indeed expected to be the most sensitive to the coexistence features appearing if there is no emergent symmetry.

Figure \ref{fig:fig6} shows results for $\Omega(g)$ at the same three temperatures for both values of $J_\perp$ as used in the histograms in
Figs.~\ref{fig:fig7} and \ref{fig:fig8}. In all cases, we see minimums close to the phase transition points in the phase diagram in Fig.~\ref{fig:fig3}.
The size dependence of the transition points as defined by the  location of the minimum is not more significant than in other definitions of the 
transition point, e.g., those shown in Fig.~\ref{fig:fig2} that we have used for the phase diagram,. The drifts may appear more pronounced here 
because we zoom in very close to the transition points. Our interest here is on the value $\Omega[g^*(L)]$ at the size-dependent locations $g^*$ 
of the minimums.

In the 2D $T=0$ system studied in Ref.~\cite{BWZhao2018}, the minimum value of the moments $I_{2q}$ decrease with increasing system size, which is the
characteristic of an emergent symmetry in the same way as in systems with ``dangerously irrelevant'' perturbations at critical points. Recent works have
used an angular order parameter $\phi_q$ similar to $\Omega$ to extract the scaling dimensions of U(1) breaking perturbations in classical \cite{Shao2020} 
and quantum \cite{Patil2021} clock models. The associated length-scale of the emergent U(1) symmetry in the ordered state can be extracted from the eventual 
increase of $\phi_q$ in the ordered phases of those models (while at their critical points the symmetry-sensitive order parameter continues to decrease as 
$L \to \infty$). Here, in Fig.~\ref{fig:fig6}(a) we see a similar effect of $\Omega(g^*)$ increasing with the system size at $T=0.1$ and $0.162$ in the 
$J_\perp=0.1$ system, thus indicating no asymptotic O(4) symmetry in accord with what we already saw in the histograms in Figs.~\ref{fig:fig7}(a) 
and \ref{fig:fig7}(b). At $T=0.168$, the minimum persists very close to $0$, but this does not indicate emergent O(4) symmetry in this case because 
the temperature is above the triple point and $\Omega(g^*)$ should trivially approach zero due to the Gaussian fluctuations.

Turning now to results for $J_{\perp}=0.01$ in Fig.~\ref{fig:fig6}(b), here $\Omega(g^*)$ is very close to $0$ and begins to increase marginally only
for the largest system sizes in each case. Thus, to within the precision of our calculations and based on the 
available system sizes (where at $T=0.06$ the largest size is $L=48$ while at the higher temperatures we have up to $L=72$, on account of the simulations 
being more time consuming at lower temperatures), these systems exhibit emergent O(4) symmetry. The length scale $\Lambda$ associated with its violation 
is at least  $\Lambda \approx 30$ at $T=0.06$ and $\Lambda \approx 60$ at $T=0.085$. These large length scales should have experimental consequences in a 
putative system with weak O(4) symmetry violation, as we will discuss further in the next section.

\section{Conclusions and Discussion}
\label{sec:v}

We have investigated a designer Hamiltonian, the layered 3D CBJQ model, which exhibits a direct quantum phase transition between PS and AFM phases when the
interlayer Heisenberg coupling $J_\perp$ is small, as expected in experimental systems where the phase transition may be realized. At $J_\perp=0$, it was previously
shown that the transition is unusual, being similar to a spin flop transition in an O(4) model \cite{BWZhao2018}. The emergent O(4) symmetry persists up to the
largest length scales studied. Here we focused on the fate of the emergent symmetry on the first-order PS-AF line up to the triple point where the system becomes
paramagnetic. For $J_\perp=0.1$, we observed the break-down of the O(4) symmetry as the system size is increased already for relatively small system sizes
$L$ of order $10$, which may already be sufficient for detectable experimental consequences, while at $J_\perp=0.01$ the order parameter retains its O(4) 
symmetry in the neighborhood of the triple point even when the length of the layers is above $50$.

Our study was largely motivated by the prospects of studying a direct PS--AF transition in the layered quantum magnet SrCu$_{2}$(BO$_{3}$)$_{2}$ 
\cite{JingGuo2020,Jimenez2020}. 
While the PS and AFM phases have been detected at high pressure and low temperatures, the pressure region where the direct transition is expected, between 2.6 and 
3 GPa, has not yet been reached at sufficiently low temperatures (likely below 1 K). We expect experiments in the near future to complete the phase diagram.

In general, phase transitions are expected to be associated with universal behaviors.
Even at a first-order transition, if the fluctuations are sufficiently large (the relevant correlation lengths sufficiently long) and originate from the proximity
of a quantum-critical point, universal behaviors should be observable up to the finite but large length scale. Given that the PS--AF transition in the CBJQ
model and SrCu$_{2}$(BO$_{3}$)$_{2}$ should both be related to the DQCP \cite{Lee2019}, we expect the results presented here to be relevant to experiments on
the latter. The exact value of the interlayer coupling $J_\perp$ is not known, and it likely depends to some extent on the pressure. Effects of $J_\perp$
in shifting the phase boundaries of the SS phases were noted in Ref.~\cite{JingGuo2020}, and the temperature dependence of $C/T$ indicates that $J_\perp/J_{2D}$
should be between $0.01$ and $0.02$ based on a comparison with weakly coupled Heisenberg layers \cite{Sengupta2003}. Our results show that the  emergent O(4)
symmetry should exist up to significant, consequential length scales with these interlayer couplings.

We have focused on $T$ close to the triple point and it is clear that the transition becomes more conventional,(and stronger 1st-order) when we reduce $T$. But it's not so clear what happens if we go to much lower $T$. If we just think of $T=0$ and increase $J_\perp$, it seems initially we should see the emergent symmetry surviving from the 2D case. So then it's not so clear if the emergent symmetry is more robust at $T=0$ or $T$ around triple point.

We here also note that the recently detected \cite{Jimenez2020} critical point in SrCu$_{2}$(BO$_{3}$)$_{2}$ at ($P=1.9$ GPa, $T = 3.3$ K) is well above the PS 
phase (which extends to about $2$ K at the same pressure). Te liquid-like dimer phase, which is still a paramagnetic phase connected to the conventional paramagnetic
phase, exists at lower pressures. Thus, the transitions into the PS and AF phases are from the conventional paramagnetic phase, and we expect our modeling of
the expected generic features of the PS--AF transition and triple point in CBJQ model to be fully relevant to SrCu$_{2}$(BO$_{3}$)$_{2}$ even though it does 
not contain the dimer phase at $T=0$, the associated ``dimer liquid'' at $T>0$ and the gas-liquid critical point.

An important issue now is how the emergent symmetry will be manifested in experiments. There should be an additional, weakly gapped Goldstone mode related to
the O(4) symmetry, in addition to the Goldstone modes arising from the O(3) symmetry of the AFM order parameter, but detailed predictions of its consequences 
for various experiments is beyond the scope of the present work. Here we only note that heat capacity
measurements between the paramagnet and the PS and AF phases in the previously inaccessible region of pressures between 2.6 and 3 GPa and at temperatures
below 1 K should be very useful. It may also be fruitful to study the grain size dependence of the low-temperature heat capacity \cite{YuanWei2020}. 
At a conventional first order phase transition the nano-scale grain thermodynamics should exhibit a finite length-scale due to the domain boundaries.
Similarly, a first-order transition with emergent continuous symmetry may exhibit effects when the grain size limits the length scale of the emergent
symmetry. 

Overall, the non-trivial transition with emergent O(4) symmetry up to large length scales provide strong motivations for the further theoretical works and experimental 
efforts on SrCu$_2$(BO$_3$)$_2$---this system so far appears to be the most promising quantum magnet for realizing ``beyond Landau'' phenomena related to the 
DQCP and unusual first-order transitions.

{\it Note added}: After the completion of the work reported
here, a DMRG study of the SS model revealed evidence for a narrow
gapless spin liquid phase between the PS and AF phases ~\cite{yang2021quantum}. Whether
or not such a phase would survive the inter-layer couplings in
SrCu$_2$(BO$_3$)$_2$ is an open question. The two competing
scenarios---the direct AF--PS transition studied here versus
a new intervening spin liquid phase---clearly motivate further
experimental and theoretical studies.

\section*{Acknowledgements}
We thank Jing Guo, Shiliang Li, Vladimir Sidorov, Liling Sun, and Ling Wang for our previous collaborations on the experimental phase diagram of
SrCu$_{2}$(BO$_{3}$)$_{2}$~\cite{JingGuo2020}, which motivated the present study. We also thank Yi-Zhuang You for useful discussions. GYU, NVM, and ZYM
acknowledge the support from the RGC of Hong Kong SAR China (Grant Nos. GRF 17303019 and 17301420), MOST through the National Key Research and Development
Program (Grant No. 2016YFA0300502) and the Strategic Priority Research Program of the Chinese Academy of Sciences (Grant No. XDB33000000). N. M acknowledge support from NSFC under Grant No.12004020. AWS was
supported by the NSF under Grant No.~DMR-1710170 and by the Simons Foundation under Simons Investigator Grant No.~511064. We thank the Center for Quantum
Simulation Sciences in the Institute of Physics, Chinese Academy of Sciences and the Tianhe-1A and Tianhe-III prototype in National Supercomputer
Center in Tianjin for their technical support and generous allocation of CPU time.

\bibliography{cbjq2019}

\end{document}